\documentclass{nature}

\usepackage[top=0.8in, bottom=1in, left=0.8in, right=0.8in]{geometry}

\usepackage{graphicx}
\usepackage{amsmath}
\usepackage{amssymb}

\newcommand{\blob}{NGC1052--DF2}
\newcommand{\msun}{M$_{\odot}$}
\newcommand{\kms}{km\,s$^{-1}$}

\title{A galaxy lacking dark matter}

\author{\large Pieter van Dokkum$^{1}$,
Shany Danieli$^{1}$,
Yotam Cohen$^{1}$,
Allison Merritt$^{1,2}$,
Aaron J.\ Romanowsky$^{3,4}$,
Roberto Abraham$^{5}$,
Jean Brodie$^{4}$,
Charlie Conroy$^{6}$,
Deborah Lokhorst$^{5}$,
Lamiya Mowla$^{1}$,
Ewan O'Sullivan$^{6}$,
Jielai Zhang$^{5}$
\vspace{8pt}}

\begin{document}

\maketitle

\begin{affiliations}
\small
 \item Astronomy Department, Yale University, New Haven, CT 06511, USA
\item Max-Planck-Institut f\"ur Astronomie, K\"onigstuhl 17, D-69117
Heidelberg, Germany
\item Department of Physics and Astronomy, San Jos\'e State University,
 San Jose, CA 95192, USA
\item University of California Observatories, 1156 High Street, Santa
Cruz, CA 95064, USA
\item Department of Astronomy \& Astrophysics, University of Toronto,
  50 St.\ George Street, Toronto, ON M5S 3H4, Canada
\item Harvard-Smithsonian Center for Astrophysics, 60 Garden Street,
Cambridge, MA, USA
 
\end{affiliations}


\begin{abstract}
Studies of galaxy surveys in the context of the cold dark matter
paradigm have shown that the mass of the dark matter halo and the total
stellar mass are coupled through a function
that varies smoothly with mass.
Their average ratio $M_{\rm halo}/M_{\rm stars}$ has a minimum of
about 30 for galaxies with stellar masses near that
of the  Milky Way (approximately $5\times 10^{10}$ solar masses)
and increases both
towards lower masses and towards higher masses.\cite{moster:10,behroozi:13b} 
The scatter in this relation is not well known; it is generally
thought to be less than a factor of two for massive galaxies
but much larger
for dwarf galaxies.\cite{more:11,oman:16}
Here we report the radial velocities of ten luminous
globular-cluster-like objects in the ultra-diffuse galaxy\cite{dokkum:15udgs}
\blob, which has a stellar mass of approximately
$2\times 10^8$ solar masses. We infer that its
velocity dispersion is less than 10.5 kilometres per
second with 90 per cent confidence, and we determine
from this that its total mass within a radius of 7.6 kiloparsecs
is less than $3.4\times 10^8$ solar masses. This implies that
the ratio $M_{\rm halo}/M_{\rm stars}$ is of order unity
(and consistent with zero), a factor of at least 400 lower than
expected.\cite{behroozi:13b}
\blob\ demonstrates that dark matter is not always coupled
with baryonic matter on galactic scales.


\end{abstract}

\blob\ was identified 
with the Dragonfly Telephoto
Array\cite{abraham:14}
in deep, wide-field imaging observations
of the NGC\,1052 group.
The galaxy is not a new discovery; it was cataloged previously
in a visual search of digitized photographic plates.\cite{karachentsev:00}
It stood out to us because of the remarkable contrast between its appearance
in Dragonfly images and Sloan Digital Sky Survey (SDSS) data:
with Dragonfly it is a low surface brightness object with some
substructure and a spatial extent of
$\sim 2'$, whereas in SDSS it appears as a collection of point-like sources.
Intrigued by the likelihood that
these compact sources are associated with the low surface
brightness object, we obtained follow-up spectroscopic
observations of \blob\
using the 10\,m W.~M.\ Keck Observatory. We also observed the galaxy with
the Hubble Space Telescope (HST).

\begin{figure*}[htb]
  \begin{center}
  \includegraphics[width=.8\linewidth]{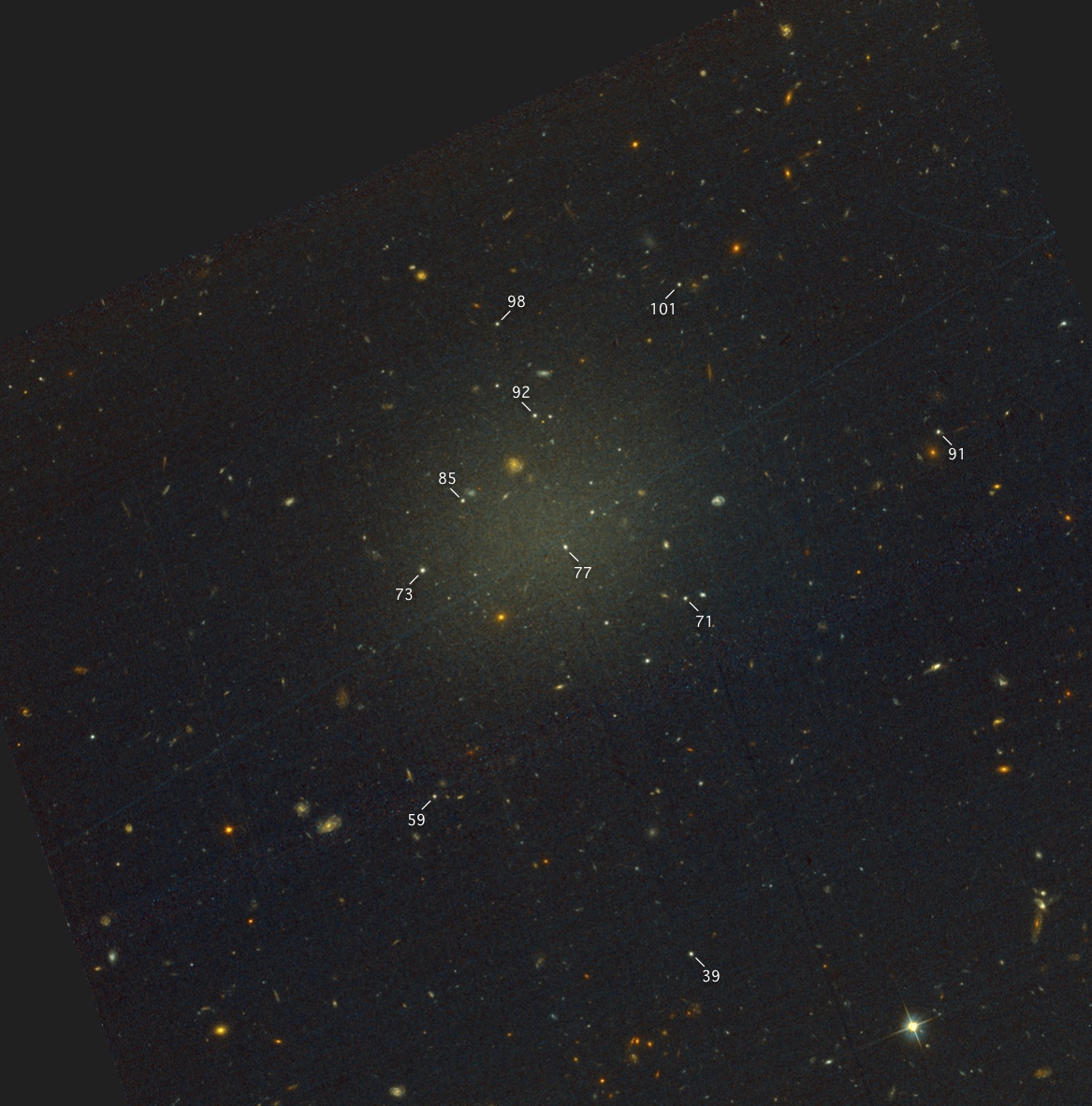}
  \end{center}
    \vspace{-0.3truecm}  
    \caption{\small \textbf{HST/ACS image of \blob.} 
\blob\ was identified as a large ($\sim 2'$) low surface object,
at $\alpha=2^h41^m46.8^s$; $\delta= -8^{\circ}24'12''$ (J2000).
Hubble Space Telescope imaging of \blob\ was obtained  2016 November 10,
using the Advanced Camera for Surveys (ACS). The exposure time was 2,180\,s
in the $V_{606}$ filter and 2,320\,s in the $I_{814}$ filter.
The image spans $3.2'\times 3.2'$, or $18.6\times 18.6$\,kpc
at the distance of \blob; North is up and East is to the left.
Faint striping is caused by imperfect CTE removal.
Ten spectroscopically-confirmed luminous compact objects are marked.
   }
   \label{hst.fig}
    \vspace{-12pt}
\end{figure*}

A color image generated from the HST $V_{606}$ and
$I_{814}$ data is shown in Fig.\ \ref{hst.fig}. The galaxy has a striking
appearance. In terms of its apparent size and surface brightness it
resembles dwarf spheroidal galaxies such as those recently
identified in the M101 group
at 7\,Mpc,\cite{danieli:17} but the fact that it is only
marginally resolved
implies that it is at a much greater distance.
Using the  $I_{814}$ band image we derive
a surface brightness fluctuation
distance of $D_{\rm SBF}=19.0\pm 1.7$\,Mpc
(see Methods).
It is located only $14'$ from the luminous
elliptical galaxy NGC\,1052, which has distance measurements
ranging from 19.4\,Mpc to 21.4\,Mpc.\cite{tonry:01,blakeslee:01fp}
We infer that \blob\ is associated with NGC\,1052,
and we adopt $D\approx 20$\,Mpc for the galaxy.


We parameterized the galaxy's structure with a two-dimensional
S\'ersic profile.\cite{galfit}
The S\'ersic index is $n=0.6$, the axis ratio is
$b/a=0.85$, the central surface brightness is
$\mu(V_{606},0) =24.4$\,mag\,arcsec$^{-2}$,
and the effective
radius along the major axis is $R_e = 22.6''$, or 2.2\,kpc. 
We conclude that \blob\ falls in the ``ultra diffuse galaxy'' (UDG)
class,\cite{dokkum:15udgs} which
have $R_e>1.5$\,kpc and $\mu(g,0)>24$\,mag\,arcsec$^{-2}$. In terms
of its structural parameters it is very similar to the galaxy
Dragonfly\,17 in the Coma cluster.\cite{dokkum:15udgs}
The total magnitude of \blob\ is $M_{606}=-15.4$, and the total
luminosity is $L_V = 1.1\times 10^8$\,L$_{\odot}$.
Its color $V_{606}-I_{814} = 0.37 \pm 0.05$ (AB), similar to that of other
UDGs and metal-poor globular clusters.\cite{dokkum:17}
The stellar mass was determined in two ways: by placing a stellar
population at $D=20$\,Mpc that matches the global properties of
\blob\ (see Methods),
and by assuming $M/L_V=2.0$ as found for
globular clusters.\cite{mclaughlin:05}
 Both methods give
$M_{\rm stars}\approx  2 \times 10^8$\,M$_{\odot}$.

\begin{figure*}[htb]
  \begin{center}
  \includegraphics[width=.92\linewidth]{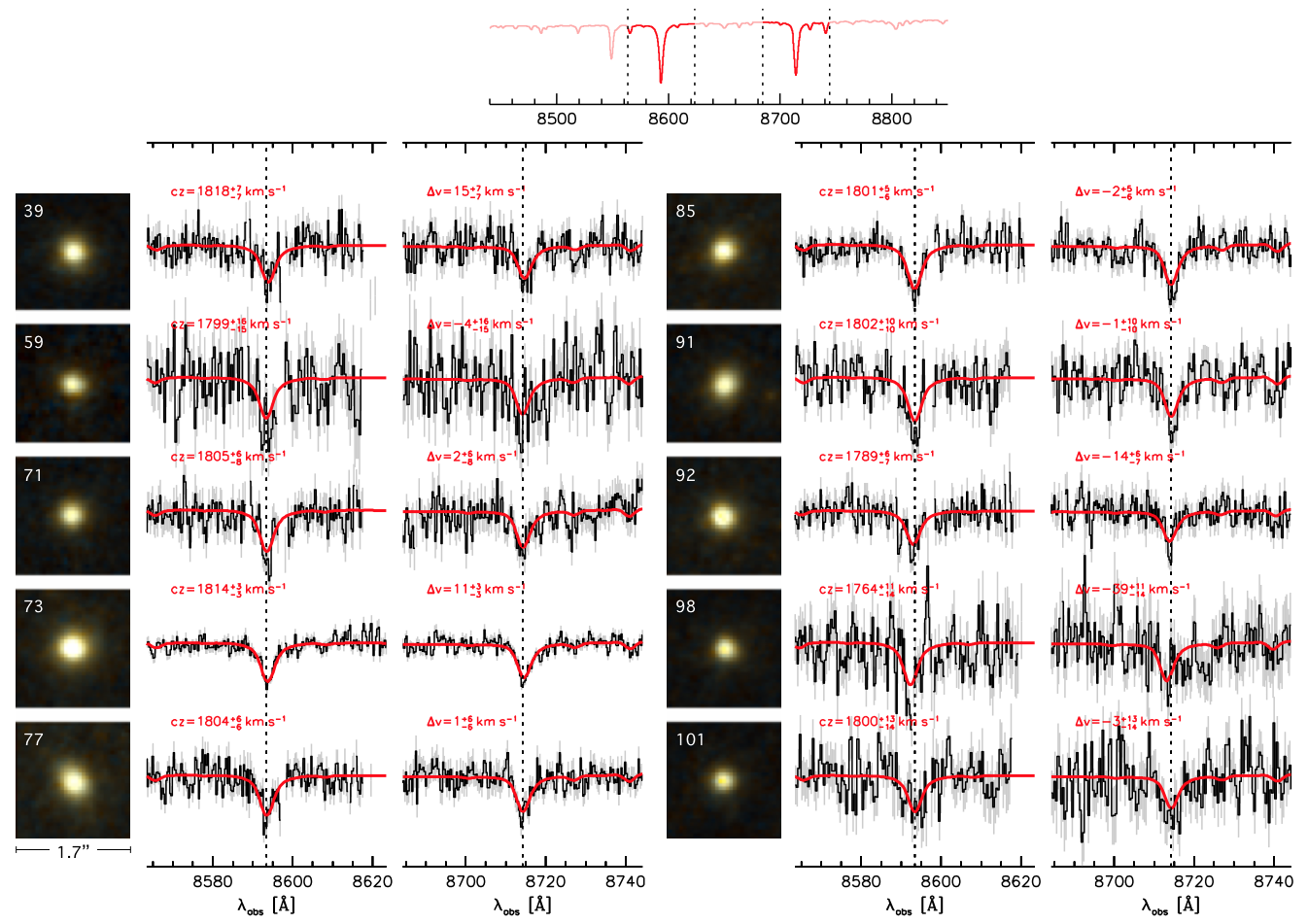}
  \end{center}
    \vspace{-0.6truecm}  
    \caption{\small \textbf{Spectra of the compact objects.}
The square panels show the HST/ACS images of the ten confirmed compact
objects. Each panel spans $1.7''\times 1.7''$, or $165\times 165$\,pc
at the distance of \blob. The Keck spectra are shown next
to the corresponding HST images.
The regions around the reddest $\lambda 8544.4$, $\lambda 8664.5$
Ca\,{\sc ii} triplet lines are shown, as illustrated
in the model spectrum at the top; the
$\lambda 8500.4$ Calcium triplet
line was included in the fit but falls on a sky line for
the radial velocity of these objects. The spectra
were obtained with LRIS, DEIMOS, or both.
The spectral resolution is $\sigma_{\rm instr} \approx 30$\,\kms.
Uncertainties in the spectra are in grey.
The signal-to-noise ratio ranges from 3.4\,pix$^{-1}$ to 12.8\,pix$^{-1}$,
with 0.4\,\AA\ pixels.
The red lines show the
best-fitting models. Radial velocities $cz$ are indicated, as well
as the velocity offset with respect to the central $\langle v \rangle
= 1803$\,\kms. This velocity is indicated with dashed vertical lines.
   }
   \label{spectra.fig}
    \vspace{-12pt}
\end{figure*}

We obtained spectroscopy of objects in the \blob\ field with
the
W.~M.\ Keck Observatory. Details of the observations and data
reduction are given in the Methods section.
We find ten objects with a radial velocity close to
1,800\,\kms\ (all other objects are Milky Way stars or background
galaxies).
We conclude that there is indeed a population of compact,
luminous objects associated with \blob.
Their spectra near the strongest
Ca triplet (CaT) lines are shown in Fig.\ \ref{spectra.fig}.
The mean velocity of the ten objects is $\langle v \rangle =
1,803^{+2}_{-2}$\,\kms.
The NGC\,1052 group has a radial velocity of $1,425$\,\kms, with
a $1\sigma$ spread of only $\pm 111$\,\kms\ (based
on 21 galaxies). \blob\
has a peculiar velocity of $+378$\,\kms\ ($3.4\sigma$) with
respect to the group, and $+293$\,\kms\ with respect to NGC\,1052
itself (Fig.\ 3). 

Images of the compact objects are shown in Fig.\ \ref{spectra.fig}
and their locations are marked on Fig.\ \ref{hst.fig}.
Their spatial distribution is somewhat more extended than that of the
smooth galaxy light: their half-number radius is $R_{\rm gc}\sim 3.1$\,kpc
(compared to $R_e = 2.2$\,kpc for the light) and the outermost object
is at $R_{\rm out}=7.6$\,kpc. In this respect, and in their compact
morphologies (they are just-resolved in our HST images, as
expected for their distance) and colors, they are similar to globular clusters
and we will refer to them as such. However,
their luminosities are much higher than those of typical globular
clusters. The brightest
(GC-73) has an absolute magnitude of
$M_{606}=-10.1$, similar to that of
the brightest globular cluster
in the Milky Way ($\omega$\,Cen). Furthermore,
the galaxy has no statistically-significant
population of
globular clusters near the canonical peak of the luminosity
function at $M_V\approx -7.5$.
The properties of these
enigmatic objects are the subject of another paper (P.v.D.\ et al.,
in preparation).

\begin{figure*}[htb]
  \begin{center}
  \includegraphics[width=.95\linewidth]{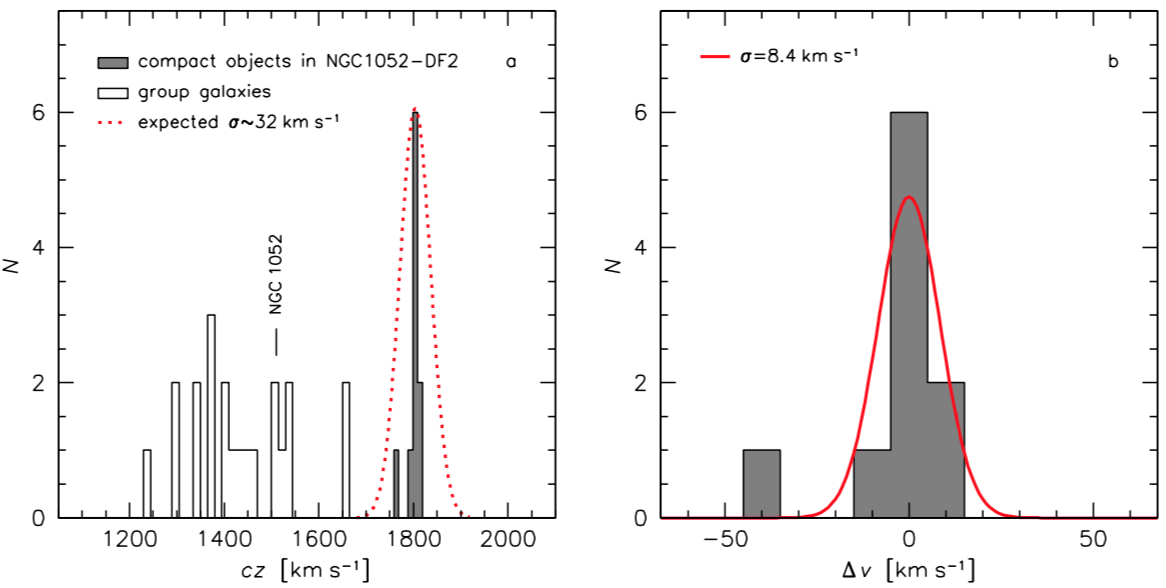}
  \end{center}
    \vspace{-0.6truecm}  
    \caption{\small \textbf{Velocity dispersion.} The filled
grey histograms
show the velocity distribution of the ten compact objects.
Panel a shows a wide velocity range, and includes the
velocities of all 21 galaxies in the NASA/IPAC Extragalactic
Database with $cz<2,500$\,\kms\ that are 
within a projected distance of two degrees from NGC\,1052.
The red dotted curve shows a Gaussian
with a width of $\sigma=32$\,\kms, the average velocity dispersion 
of Local Group galaxies with $8.0\leq \log(M_{\rm stars}/{\rm M}_{\odot})
\leq 8.6$.
Panel b shows a narrow velocity range centered on
$cz=1,803$\,\kms. 
The red solid curve is a Gaussian with a width that is equal to the 
biweight dispersion of the velocity
distribution of the compact objects, $\sigma_{\rm obs}=8.4$\,\kms.
Taking observational errors into account, we derive an intrinsic
dispersion of $\sigma_{\rm intr}=3.2^{+5.5}_{-3.2}$\,\kms.
The 90\,\% confidence upper limit on the intrinsic dispersion
is $\sigma_{\rm intr}<10.5$\,\kms.
   }
   \label{vhist.fig}
    \vspace{-6pt}
\end{figure*}

\begin{figure*}[htb]
  \begin{center}
  \includegraphics[width=.95\linewidth]{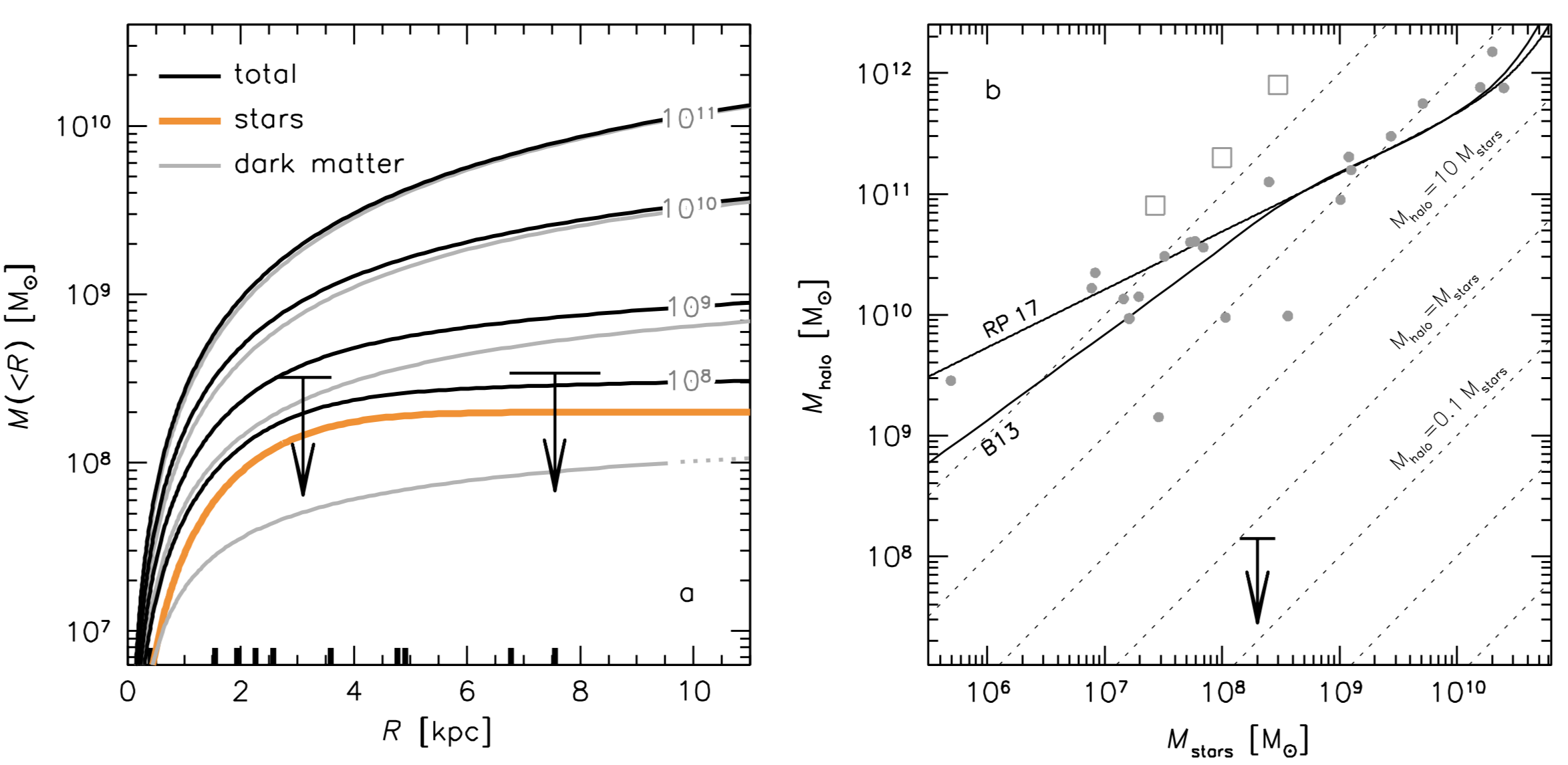}
  \end{center}
    \vspace{-0.6truecm}  
    \caption{\small \textbf{Constraints on the halo mass.}
Panel a shows enclosed mass profiles for NFW halos\cite{navarro:97}
of masses $M_{200}=10^8$\,\msun, $10^9$\,\msun, 
$10^{10}$\,\msun, and $10^{11}$\,\msun\ (grey lines). The $10^8$\,\msun\
halo profile is shown by a dotted line beyond
$R=R_{200} \sim 10$\,kpc. The orange curve is the enclosed mass
profile for the stellar component, and the black curves are the total
mass profiles $M_{\rm total} = M_{\rm stars} + M_{\rm halo}$.
The ten globular clusters are at distances ranging from $R=0.4$\,kpc to
$R=7.6$\,kpc; short
vertical bars on the horizontal axis
indicate the locations of individual clusters.
The 90\,\% upper limits on the total enclosed mass of \blob\
are shown by arrows. The limit at $R=7.6$\,kpc was determined
with the TME method.\cite{watkins:10} The arrow at $R=3.1$\,kpc is 
the mass limit
within the half-number radius of the globular cluster system.\cite{wolf:10}
The dynamical mass of \blob\ is consistent with
the stellar mass, and leaves little room for a dark matter halo.
Panel b shows the upper limit on the halo mass, and compares
this to the expected dark matter mass from studies that model
the halo mass function and the evolution of
galaxies.\cite{behroozi:13b,rodriguez:17} Grey solid symbols are nearby dwarf
galaxies with rotation curves extending to at least two disk
scale lengths.\cite{oman:16} Open squares are three cluster UDGs with measured
kinematics: VCC\,1287,\cite{beasley:16} Dragonfly\,44,\cite{dokkum:17} and
DFX1.\cite{dokkum:17}
\blob\ falls a factor of $\gtrsim 400$
below the canonical relations.
   }
   \label{halo.fig}
    \vspace{-12pt}
\end{figure*}

The central observational result of the present study is the remarkably
small spread among the velocities of the ten clusters (Fig.\
\ref{vhist.fig}). The observed velocity dispersion is $\sigma_{\rm obs}
=8.4$\,\kms, as measured with the biweight estimator (see Methods). This value
is much smaller than that in previously studied (cluster)
UDGs,\cite{beasley:16,dokkum:17}
and not much higher than the expectation from observational
errors alone. Taking the errors into account, we find an
intrinsic dispersion of $\sigma_{\rm intr} = 3.2^{+5.5}_{-3.2}$\,\kms.
The 90\,\% confidence
upper limit is $\sigma_{\rm intr}<10.5$\,\kms.
To our knowledge
this is the coolest known galaxy outside of the Local Group.
Within the Local Group, 
typical galaxies
with velocity dispersions in this range
are small ($R_e
\sim 200$\,pc) and have a
low stellar mass ($M_{\rm stars}\sim 2-3\times
10^6$\,\msun).\cite{mcconnachie:12}
The average velocity dispersion of Local Group galaxies
with $8.0 \leq \log(M/{\rm M}_{\odot}) \leq 8.6$ is 32\,\kms\
(dotted curve in Fig.\ \ref{vhist.fig}a).

We calculate the corresponding 90\,\% confidence upper limit
on the mass of \blob\ using the tracer mass estimator (TME)
method,\cite{watkins:10} which provides an estimate of the
dynamical mass within the radius of the outermost discrete tracer.
We find $M_{\rm dyn} < 3.4 \times 10^8$\,\msun\
within $R_{\rm out}=7.6$\,kpc. 
We also determine the
dynamical mass within the projected half-number radius of the globular
cluster system,\cite{wolf:10} and find $M_{\rm dyn}<3.2\times 10^8$\,\msun\
within $R_{\rm gc}=3.1$\,kpc.

In 
Fig.\ \ref{halo.fig}a the enclosed mass is compared to the expected
mass from the stars alone (orange line) and to
models with different halo masses. The dynamical mass is
consistent with the stellar mass, leaving little room for dark
matter.  The best
fit to the kinematics is obtained for $M_{\rm halo}=0$, and 
the $90$\,\% confidence upper limit on the dark matter halo mass
is $M_{\rm halo}< 1.5\times 10^8$\,\msun.
We note that the combination of the large spatial extent and
low dynamical mass of \blob\ yields an unusually robust
constraint on the total halo mass. Typically, kinematic tracers are
only available out to a small fraction of $R_{200}$, and
a large extrapolation is required to convert the measured enclosed mass to
a total halo mass.\cite{oman:16} However,
for a halo of mass $M_{200}\sim 10^8$\,\msun\
the virial radius is only $\sim 10$\,kpc,  similar to the radius
where the outermost globular clusters reside.
As shown in Fig.\ \ref{halo.fig}b,
a galaxy with a stellar mass of $M_{\rm stars}=2\times 10^8$\,\msun\
is expected to have a halo mass
of $M_{\rm halo} \approx 6\times 10^{10}$\,\msun, a factor of $\sim 400$
higher than the upper limit that we derive.
We conclude that \blob\ is extremely
deficient in dark matter, and a good candidate for
a ``baryonic galaxy'' with no dark matter at all.

It is unknown how the galaxy was formed. One possibility is that
it is an old tidal dwarf, formed from gas that was flung out of
merging galaxies.
Its location near an elliptical galaxy and its high peculiar
velocity are consistent with this idea. 
Its relatively
blue color suggests a lower metallicity than might be expected
for such objects,\cite{bournaud:07} but that depends on the
detailed circumstances of its formation.\cite{ploeckinger:18}
An alternative explanation is that the galaxy formed
from low metallicity
gas that was swept up in quasar winds.\cite{natarajan:98}
The lack of dark matter, the
location near a massive elliptical, the peculiar velocity, and the color
are all qualitatively consistent with this scenario, although it is
not clear whether the large
size and low surface brightness of \blob\ could have been produced by this
process. A third option is that the galaxy formed from inflowing gas
that fragmented before reaching NGC\,1052, either
relatively close to the assembling
galaxy\cite{canning:14} or out in the halo.\cite{mandelker:17}
This fragmentation may have been aided or precipitated by jet-induced
shocks.\cite{vanbreugel:85}
In any scenario the luminous globular cluster-like objects require
an explanation; generically, it seems likely that the three
peculiar aspects of the galaxy (its large size, its low dark matter
content, and its population of luminous compact objects) are related.
An important missing piece of information is the number density
of galaxies such as \blob.
There are several other objects in our Cycle 24
HST program that look broadly similar, but these
do not have dynamical measurements yet -- and the fact that other
UDGs have anomalously {\em high} dark matter
fractions\cite{beasley:16,dokkum:17}
demonstrates that such data are needed to interpret these galaxies.

Regardless of the formation history of \blob, its existence
has implications for the dark matter paradigm.
Our results demonstrate that
dark matter  is separable from galaxies, which is (under certain
circumstances) expected if it is bound to baryons through
nothing but gravity.
The
``bullet cluster'' demonstrates that dark matter does not always
trace the bulk of the
baryonic mass,\cite{clowe:06} which in clusters is in the form of gas.
\blob\ enables us to make the complementary point
that dark matter does not always coincide with galaxies either:
it is a distinct ``substance'' that may or may not be present
in a galaxy. Furthermore, and 
paradoxically, the existence of \blob\ may falsify alternatives
to dark matter. In theories
such as MOND\cite{milgrom:83} and
the recently proposed emergent gravity paradigm\cite{verlinde:16}
a ``dark matter'' signature  should {\em always} be detected,
as it is an unavoidable
consequence of the presence of ordinary matter.
In fact, it had been argued
previously\cite{kroupa:12} that the apparent
{\em absence} of galaxies such as \blob\ constituted a falsification
of the standard cosmological model, and evidence for modified
gravity.
For a MOND acceleration scale
of
$a_0= 3.7\times 10^3$\,km$^2$\,s$^{-2}$\,kpc$^{-1}$ the
expected\cite{angus:08} velocity dispersion of \blob\ is
$\sigma_{\rm M} \approx \left( 0.05\,GM_{\rm stars} a_0 \right)^{1/4}
\approx 20$\,\kms, a factor of two higher than the 90\,\% upper
limit on the observed dispersion.

\vspace{0.3cm}
\noindent
{\bf METHODS}\vspace{0.2cm}\\
{\bf Imaging.} In this paper we use imaging 
from the Dragonfly Telephoto Array, the Sloan Digital
Sky Survey,  the MMT, the Gemini
North telescope, and the Hubble Space Telescope.
\vspace{0.1cm}\\
{\it Dragonfly:} The Dragonfly Telephoto Array\cite{abraham:14} data 
were obtained in the context of the Dragonfly Nearby Galaxy
Survey.\cite{merritt:16a}  Dragonfly  was operating with 8
telephoto lenses at the time of the observations, forming the optical
equivalent of an f/1.0 refractor with a 40\,cm aperture.
The data reach a $1\sigma$ surface brightness limit of
$\mu(g) \sim 29$\,mag\,arcsec$^{-2}$ in $12''\times 12''$
boxes.\cite{merritt:16a}  The full Dragonfly field is
shown in \ref{dragonfly.fig}, as well as the area
around NGC\,1052 and \blob.
\vspace{0.1cm}\\
{\it SDSS:}  SDSS images in $g$, $r$, and
$i$ were obtained from the DR14 Sky Server.\cite{sdss14}
To generate the image in Extended Data Figure 4
the data in the three bands were
summed without weighting or scaling.
The object is located near the corner of a frame.
We note that the SDSS photometry for the compact objects is not
reliable, as it is ``contaminated'' by the low surface brightness
 emission of the galaxy (which is
just detected in SDSS).
\vspace{0.1cm}\\
{\it MMT:} MMT/Megacam imaging of the NGC\,1052 field was available
from a project to image the globular cluster systems of nearby
early-type galaxies.\cite{napolitano:09} 
The data were taken in the $r$ and $i$ bands, in $0.9\arcsec$
seeing. They were used for target selection in our first Keck
spectroscopic run.
\vspace{0.1cm}\\
{\it Gemini:} 
We obtained imaging with the
Gemini-North Multi Object Spectrograph\cite{gmos} (GMOS) in
program GN-2016B-DD-3. The observations were executed on 2016 October 10,
with total exposure times of 3,000\,s in the $g$ band
and 3,000\,s in the $i$ band. The seeing was $0.65''$ in $i$ and
$0.70''$ in $g$. The data were reduced using the Gemini
IRAF package.  Low order
polynomials were fitted to individual (dithered) 300\,s frames after
carefully masking objects, 
to reduce large scale background
gradients at low surface brightness levels.
The images were used to aid in the target
selection and mask design for the LRIS spectroscopy.
The Gemini data also provide the best available information
on the regularity of the galaxy at low surface brightness levels
(see below).
The combined frames still show some background artifacts
but they are less prominent than those in the HST data.
Finally, a visual inspection of the
Gemini images prompted us to request a change in the scheduling of
HST program GO-14644, moving the already-planned ACS observation of \blob\ 
to an earlier date. 
\vspace{0.1cm}\\
{\it HST:} The HST data were obtained as part of program GO-14644. The
aim of this program is to obtain ACS images of
a sample of 23 low surface brightness objects that were identified in
fields of the Dragonfly Nearby Galaxy Survey.\cite{merritt:16a} 
\blob\ was observed on 2016 November 16, for a total of two orbits.
Exposure times were 2,180\,s in $V_{606}$ and 2,320\,s in $I_{814}$.
In this paper we use drizzled .drc images, which
have been corrected for charge transfer efficiency (CTE) effects.
Despite this correction some CTE artifacts are still visible in the data
(see Fig.\ 1).
\vspace{0.2cm}\\
{\bf Structural parameters.} The size, surface brightness, and
other structural parameters of \blob\ were determined from the HST data.
First, the $I_{814}$ image was rebinned to a lower
resolution to increase the S/N ratio per pixel. Then, a preliminary
object mask was created from a segmentation map
produced by SExtractor,\cite{bertin:96}
using a relatively high
detection threshold.
A first-pass S\'ersic model\cite{sersic} for the galaxy was obtained using the
GALFIT software.\cite{galfit} This model was subtracted from the
data, and an improved object mask was created using a lower SExtractor
detection threshold. Finally, GALFIT was run again to obtain the final
structural parameters and total magnitude.
The total magnitude in the $V_{606}$ band, and the
$V_{606}-I_{814}$ color, were determined by running GALFIT on the (binned
and masked) $V_{606}$
image with all parameters except the total
magnitude fixed to the $I_{814}$ values. The structural parameters,
total magnitude, and color are listed in the main text.
We note that we measured nearly identical structural parameters from the
Gemini images.
\vspace{0.2cm}\\
{\bf Spectroscopy.} We obtained spectroscopy of compact objects in the
\blob\ field in two observing runs. The first set of observations
was obtained on 2016 September 28 with
the Deep Imaging Multi-Object
Spectrograph\cite{deimos} (DEIMOS) on Keck II, and the second
was obtained
on 2016 October 26 and 27 
with the Low-Resolution Imaging Spectrometer\cite{oke:95} (LRIS) on
Keck I.\vspace{0.1cm}\\
{\it DEIMOS observations:} Conditions were variable,
with cirrus clouds increasing throughout the night.
We obtained 4\,hrs of total on-source
exposure time on a single multi-object mask; a second mask
was exposed but yielded no useful data due to clouds.
The target selection algorithm gave priority
to compact objects with $i\lesssim 22.5$
near \blob, selected from the MMT data.
We used the 1,200\,lines\,mm$^{-1}$ grating
with a slit width of $0.75''$, providing an instrumental
resolution of $\sigma_{\rm instr}\approx 25$\,\kms.
The data were reduced with the same pipeline that we used
previously\cite{dokkum:17,dokkum:16} for the Coma UDGs Dragonfly~44
and DFX1. The globular clusters
GC-39, GC-71, GC-73, GC-77, GC-85, GC-92, and GC-98 (see Fig.\ 1 and 2)
were included in this mask.
\vspace{0.1cm}\\
{\it LRIS observations:} 
Two multi-slit masks were observed, one for 3.5 hrs (on source)
on October 26 and a second for 4 hrs on October 27. Targets were
selected from the Gemini data, giving priority to compact objects
that had not been observed with DEIMOS.
Conditions were fair during both nights, with intermittent cirrus and seeing
of $\approx 1''$.
We used the 1,200\,lines\,mm$^{-1}$ gold coated
grating blazed at 9,000\,\AA. The instrumental resolution
$\sigma_{\rm instr}\approx 30$\,\kms. A custom pipeline was used
for the data reduction, modeled on the one that we developed for
DEIMOS. LRIS suffers from significant
flexure, and  the main difference with the DEIMOS pipeline is that each
individual 1800\,s exposure was reduced and calibrated independently to avoid
smoothing of the combined spectra in the wavelength direction.
The clusters GC-39, GC-59, GC-73, GC-91, and GC-101
were included in the first mask; GC-39, GC-71, GC-77, GC-85, and GC-92
were included in the second.
\vspace{0.1cm}\\
{\it Combined spectra:} Most compact objects were observed multiple times,
and we combined these individual spectra to increase the S/N ratio. 
All spectra were given equal weight, and prior to combining they
were divided by a low order polynomial fit to the continuum in the
CaT region. We tested that weighting by the
formal S/N ratio instead does not change the best-fit velocities.
The individual spectra were also shifted in
wavelength to account for the heliocentric correction; this needs
to be done at this stage as the DEIMOS and LRIS data were taken
one month apart. Six GCs have at least two independent
observations and effective exposure times of $\approx 8$\,hrs;
four were observed only once: GC-59, GC-91, GC-98, and GC-101.
The S/N ratio in the final spectra ranges from 3--4\,pix$^{-1}$ for
GC-59, GC-98, and GC-101
to 13\,pix$^{-1}$ for GC-73. A pixel is
0.4\,\AA, or 14\,\kms.
\vspace{0.2cm}\\
{\bf Velocity measurements.} Radial velocities were
determined for all objects with detected
CaT absorption lines. No fits were attempted for
background galaxies (based on the detection of redshifted emission
lines), Milky Way stars, or spectra with no visible features.
The measurements were performed 
by fitting a template spectrum to the
observations, using the  {\tt emcee} 
MCMC algorithm.\cite{emcee} 
The template is a high resolution stellar population synthesis
model,\cite{conroy:09}
smoothed to the instrumental resolution. The model has an age of
11\,Gyr and a metallicity of [Fe/H]$=-1$, which is consistent
with the colors of the compact objects ($V_{606}-I_{814}\sim 0.35$);
the results are independent of the precise choice of template.
The fits are performed over the observed wavelength range
$8,530\,{\rm \AA}\leq \lambda \leq 8,750\,{\rm \AA}$ and have three
free parameters: the radial velocity, the normalization, and
an additive term that serves as a template mismatch parameter (as
it allows the strength of the absorption lines to vary with respect
to the continuum).
The fit was performed twice. After the first pass all pixels that
deviate $>3\sigma$ are masked in the second fit. This step reduces
the effect of systematic sky subtraction residuals on the fit.

The uncertainties given by the {\tt emcee} method do not take
systematic errors into account.
Following
previous studies\cite{dokkum:09}
we determined the
uncertainties in the velocities by shuffling the residuals.
For
each spectrum, the best-fitting
model was subtracted from the data. Next, 500 realizations of the
data were created by randomizing the wavelengths of the residual
spectra and then adding the shuffled residuals to the
best-fitting model. These 500 spectra were then fit using a simple
$\chi^2$ minimization, and the 16$^{\rm th}$ and 84$^{\rm th}$ percentiles
of the resulting velocity distribution yield the error bars.
In order to preserve the higher noise at the location of sky lines
the randomization was done separately for pixels at the locations
of sky lines and for pixels in between the lines.
We find that the resulting errors show the expected inverse trend
with the S/N ratio of the spectra, whereas the {\tt emcee} errors
show large variation at fixed S/N ratio.

We tested the reliability of the errors by applying the same procedure
to the individual LRIS and DEIMOS spectra for the six objects that were
observed with both instruments
(GC-39, GC-71, GC-73, GC-77, GC-85, and GC-92).
For each object
the observed difference between the LRIS and DEIMOS velocities was divided by
the expected error in the difference. The rms of these ratios is $1.2\pm 0.3$,
that is, the empirically-determined uncertainties
are consistent with the observed differences
between the independently-measured LRIS and DEIMOS velocities.
\vspace{0.2cm}\\
{\bf Velocity dispersion.}
The observed
velocity distribution of the ten clusters is not well approximated by
a Gaussian. Six of the ten have velocities that are within $\pm 4$\,\kms\
of the mean and one is $39$\,\kms\ removed from the mean.
As a result, different ways to estimate the Gaussian-equivalent
velocity spread $\sigma_{\rm obs}$ yield
different answers. The normalized median absolute deviation
$\sigma_{\rm obs,nmad} = 4.7$\,\kms, the biweight\cite{beers:90}
$\sigma_{\rm obs,bi} = 8.4$\,\kms, and the rms $\sigma_{\rm obs,rms}
=14.3$\,\kms. The rms is driven by one object with a relatively
large velocity uncertainty (GC-98), and is
inconsistent with the velocity distribution of the other nine.
Specifically, for 10 objects drawn from a Gaussian distribution
and including the observed errors
the probability of measuring $\sigma_{\rm bi}\leq 8.4$
if $\sigma_{\rm rms}\geq 14.3$  is 1.5\,\%,
and the probability of measuring $\sigma_{\rm nmad}\leq 4.7$ is
$3\times 10^{-3}$. We therefore adopt the biweight dispersion
rather than the rms when determining the intrinsic dispersion
below. We then show that the
presence of GC-98 is consistent with the intrinsic dispersion
that we derive using this statistic.

The observed dispersion has to be corrected for
observational errors, which are of the same order as $\sigma_{\rm obs}$
itself. We determined the intrinsic dispersion and its uncertainty
in the following way. For a given value of $\sigma_{\rm test}$
we generated 1000 samples of 10 velocities, distributed according
to a Gaussian of width $\sigma_{\rm test}$. The ten velocities in
each sample were then perturbed with errors, drawn from Gaussians with
widths equal to the empirically-determined uncertainties in the
measured dispersions. Using the biweight estimator, ``measured''
dispersions $\sigma_{\rm obs,test}$ were calculated for all samples. If
the value 8.4 is within the $16^{\rm th}$ -- $84^{\rm th}$
percentile of the distribution of $\sigma_{\rm obs,test}$
then $\sigma_{\rm test}$ is
within the $\pm 1\sigma$ uncertainty on $\sigma_{\rm intr}$.
This method gives $\sigma_{\rm intr} = 3.2^{+5.5}_{-3.2}$\,\kms.
As the intrinsic dispersion is consistent with zero, a more
meaningful number than the best-fit is the 90\,\%
confidence upper limit; we find
$\sigma_{\rm intr}<10.5$\,\kms.

We now return to the question whether GC-98 is consistent with the other
nine objects. This cluster has a velocity offset of $\Delta v =
-39^{+11}_{-14}$\,\kms. For the upper limit on the intrinsic
dispersion ($\sigma_{\rm intr}=10.5$\,\kms) the object 
is a $2.4\sigma$ outlier, and the probability of having a $>2.4\sigma$
outlier in a sample of ten is 15\,\%.
Interestingly the combination of the biweight
constraint of $\sigma_{\rm intr}<10.5$ and the existence of GC-98
implies a fairly narrow range of intrinsic dispersions that are
consistent with the entire set of ten velocities (assuming that they are drawn
from a Gaussian distribution and the errors are correct).
The probability of having at least one object with the velocity
of GC-98 is $<10$\,\% if the intrinsic dispersion is $\sigma_{\rm intr}<8.8$\,\kms.
Taking both 90\,\% confidence limits at face value, the allowed range in
the intrinsic dispersion is $8.8$\,\kms\,$<\sigma_{\rm intr}<10.5$\,\kms.
\vspace{0.2cm}\\
{\bf Expected dispersion from Local Group galaxies.}
In Fig.\ \ref{vhist.fig}a
we illustrate how unusual the kinematics
of \blob\ are by comparing the observed velocity distribution
to that expected from
Local Group galaxies with the
same stellar mass (broken red curve). The width of this Gaussian was calculated
from the SEPT2015 version of the Nearby Dwarf Galaxies
catalog.\cite{mcconnachie:12}  The catalog has entries for
both velocity
dispersions and rotation velocities, and for both gas and stars.
To obtain a homogeneous estimate we use ``effective'' dispersions,
$\sigma_{\rm eff} \equiv                         
(\sigma^2 + 0.5 v_{\rm rot}^2)^{0.5}$. When both gas and
stellar kinematics are available we use the highest value
of $\sigma_{\rm eff}$, as this typically is a rotation curve
measurement from gas at large radii
versus a stellar velocity dispersion
measurement. Stellar masses were calculated directly
from the $V$ band absolute magnitude assuming $M/L_V=2.0$, for
consistency with \blob. Five galaxies have
a stellar mass that is within a factor of two of that of
\blob: IC~1613, NGC~6822, Sextans~B, and the M31 satellites
NGC~147 and NGC~185. The average dispersion
of these five galaxies is $\langle \sigma_{\rm eff}\rangle = 32$\,\kms,
with an rms variation of 8\,\kms.

In \ref{local.fig} we compare \blob\ to the nearby dwarf sample
in the plane of velocity dispersion versus half-light radius,
with the size of the symbols indicating the stellar mass.
Comparing \blob\ to other galaxies with velocity dispersions
in this range,
we find that its size is larger by a factor of $\sim 10$ and its stellar
mass is larger by a factor of $\sim 100$.
\vspace{0.2cm}\\
{\bf Distance.} 
The heliocentric radial velocity of \blob\ is $1,803\pm 2$\,\kms, or
$1,748\pm 16$\,\kms\
after correcting for the effects of the Virgo cluster, the great
attractor, and the Shapley supercluster on the local velocity
field.\cite{mould:00} For $H_0 = 70\pm 3$\,\kms\,Mpc$^{-1}$ a
Hubble flow distance
of $D_{\rm HF}=25\pm 1$\,Mpc is obtained.
However, the proximity to NGC\,1052 
($14'$, or $\approx 80$\,kpc in projection) strongly suggests that
\blob\ is associated with this massive elliptical galaxy. 
The distance to NGC\,1052, as determined from surface brightness
fluctuations and the fundamental plane,\cite{tonry:01,blakeslee:01fp} is
$D_{\rm 1052}=20.4 \pm 1.0$\,Mpc. The velocity of \blob\ with
respect to NGC\,1052 is then $+293$\,\kms.

A third distance estimate can be obtained
from the luminosity function of the
compact objects. As discussed in another paper (P.v.D.\ et al.,
in preparation) the luminosity function has a narrow peak at $m_V
\approx 22.0$. The canonical globular cluster luminosity function
can be approximated by a Gaussian with
a well-defined peak\cite{rejkuba:12} at $M_V\approx -7.5$.
If the compact objects are typical globular clusters, the implied distance
is $D_{\rm GC} \approx 8$\,Mpc.
This is an important possibility, as
the main conclusions of the paper would be 
weakened considerably if the galaxy is
so close to us. For this distance the stellar mass estimate is
an order of magnitude lower:
$M_{\rm stars} \approx 3\times 10^7$\,\msun. The four
Local Group galaxies that have a stellar mass within a factor of
two of this value (Fornax, Andromeda\,II, Andromeda\,VII, and UGC\,4879)
have a mean dispersion of $\langle \sigma_{\rm eff}\rangle =
11.7\pm 0.5$\,\kms, only slightly higher than the
upper limit to the dispersion
of \blob. The peculiar velocity of the galaxy would be $\sim
1200$\,\kms; this is of course extreme, but it is difficult to argue
that it is less likely than having a highly peculiar globular
cluster population and a lack of dark matter.

Fortunately, we have independent information to verify the distance,
namely the appearance of \blob\ in the HST images.
In \ref{sbf.fig}a we show the
central $33\arcsec \times 33\arcsec$ of the galaxy in the HST $I_{814}$
band. A smooth model of the galaxy, obtained by median filtering
the image, was subtracted; background galaxies and globular clusters
were masked. The
mottled appearance is not due to noise but due to the variation
in the number of giants contributing to each pixel.
Following previous studies,\cite{tonry:01,mei:05,blakeslee:10}
we measure the surface brightness fluctuation (SBF) signal from
this image and determine
the distance from the SBF magnitude.

The  azimuthally-averaged power spectrum of the image is shown in
\ref{sbf.fig}b. As is customary\cite{mei:05} the smallest and largest
wavenumbers are omitted, as they are dominated by, respectively,
residual large
scale structure in the image and noise correlations. 
Again following previous studies,\cite{mei:05,blakeslee:10} the
power spectrum is
fit by a combination of a constant (dotted line)
and the expectation power spectrum $E(k)$ (dashed line). The expectation
power spectrum is the convolution
of the power spectrum of the PSF and that of the window function.
The window function is the square root
of the median-filtered model of the galaxy, multiplied by the mask containing
the globular clusters and background galaxies.

The normalization of $E(k)$ is the SBF magnitude, $\overline{m}_{814}$.
We find $\overline{m}_{814} = 29.45\pm 0.10$. Using Eq.\ 2 in
ref.\ 47, $V_{606}-I_{814}=0.37 \pm 0.05$,
and $g_{475}-I_{814} = 1.852 (V_{606}-I_{814}) + 0.096$,
the  absolute SBF magnitude is
$M_{814} = -1.94\pm 0.17$. The uncertainty is a combination of the
error in the $V_{606}-I_{814}$ color and the systematic uncertainty
in the extrapolation of the relation between $g_{475}-I_{814}$ color
and $M_{814}$ (as determined from the difference between Eqs.\ 1
and 2 in ref.\ 47).
The SBF distance modulus $\overline{m}-\overline{M}
= 31.39 \pm 0.20$,
and the SBF distance $D_{\rm SBF}
= 19.0 \pm 1.7$\,Mpc. This result is consistent
with $D_{1052}$, and rules out the ``globular cluster distance'' of
$D_{\rm GC}=8$\,Mpc.
\vspace{0.2cm}\\
{\bf Stellar mass.} 
We determined the stellar mass from a stellar population synthesis
model.\cite{conroy:09}
A two-dimensional model galaxy was created using the
ArtPop code\cite{danieli:18}
that matches the morphology, luminosity, color, and SBF signal of \blob.
The model has
a metallicity [Z\,/\,H]$\,\,=-1$ and an age of 11\,Gyr.
These parameters
are consistent
with the regular morphology of the galaxy and
with spectroscopic constraints on the stellar populations
of Coma cluster UDGs.\cite{gu:17}
For a Kroupa IMF\cite{kroupa:01}
the stellar mass of this model is
$1.8\times 10^{8}$\,\msun, similar to that obtained from a simple 
$M/L_V=2.0$ conversion\cite{mclaughlin:05} ($M_{\rm stars}=
2.2\times 10^8$\,\msun). In the main text and below 
we assume $M_{\rm stars}\approx 2 \times 10^{8}$\,\msun.
We note
that the uncertainty in the stellar mass is much smaller than that in the
dynamical mass, for reasonable choices of the IMF.
\vspace{0.2cm}\\
{\bf Dynamical equilibrium.} Some large low surface brightness
objects are almost certainly in the process of disruption; examples are the
``star pile'' in the galaxy cluster Abell 545,\cite{struble:88,salinas:11}
the boomerang-shaped galaxy DF4 in the field of M101,\cite{merritt:16}
and And\,XIX in the Local Group\cite{collins:13} (marked in 
\ref{local.fig}).  And\,XIX, and also And\,XXI and And\,XXV, are
particularly informative as they combine large sizes
with low velocity dispersions,\cite{tollerud:12} and it has been suggested that
tidal interactions have contributed to their unusal
properties.\cite{collins:13} (We note that these galaxies
are not direct analogs of \blob: the stellar masses of
these Andromeda satellites are a factor of $\sim 100$ lower than that
of \blob, and their dynamical $M/L$ ratios are at least
a factor of $10$ higher.)
In \ref{gemini.fig}c we show the Gemini $i$ band image of
\blob\ (along with the Dragonfly and SDSS images).
There is no convincing
evidence of strong position angle twists
or tidal features at least out to $R\sim 2 R_e$
(see also the Dragonfly image in panel a).
The regular appearance strongly suggests 
that the object  has survived in its present form for multiple
dynamical times, and we infer that the kinematics can likely
be interpreted in the context of a system that is in dynamical
equilibrium.

We note that
the regular morphology of \blob\ also
provides an interesting constraint on its
formation time: the orbital velocity in the outer parts is $\gtrsim 5$\,Gyr,
which means it has to have formed very early in order to lose any
sign of its assembly. Furthermore,  it
provides a lower limit for the 3D distance between \blob\ and
the massive elliptical galaxy NGC\,1052.
The Jacobi radius (i.e., the distance from the center
of the galaxy to the first Lagrangian point) is given by\cite{baumgardt:10}
\begin{equation}
R_J = \left( \frac{GM}{2V_{1052}^2}\right)^{1/3} R_{1052}^{2/3},
\end{equation}
with $R_{1052}$ the distance between \blob\ and
NGC\,1052 and $V_{1052}$ the circular velocity of NGC\,1052.
Taking $R_J>5$\,kpc, $M\sim
2\times 10^{8}$\,\msun, and
$V_{1052}\sim 300$\,\kms\ (the velocity difference
between the two galaxies, as well as $\sim \sqrt{2}
\sigma_{1052}$), we obtain $R_{1052} \gtrsim 160$\,kpc,
a factor of two larger than the projected distance.
\vspace{0.2cm}\\
{\bf Source of dynamical support.}
The morphology of the galaxy strongly indicates that it is supported
by random motions rather than rotation: the S\'ersic index is 0.6,
similar to that of dSph galaxies; the isophotes are elliptical rather
than disky; and there are no bars, spiral arms, or other features
that might be expected in a thin disk. The galaxy has not been
detected in moderately deep
H\,{\sc i} observations.\cite{mckay:04} It is also difficult to
imagine a physical model for the formation of a huge, extremely
thin disk of massive blue globular clusters, even in
spiral galaxies: although the kinematics
of the metal-rich
globular cluster population in M31 are clearly related to its disk,
the metal-poor ones have a large velocity dispersion.\cite{caldwell:16}
Finally, there is no evidence for a velocity gradient.
In \ref{rotation.fig}a we show the measured
velocities of the globular clusters as a function of the projected
distance along the major axis. There is no coherent pattern.
Based on these arguments
our default mass measurement assumes that the galaxy is
supported by random motions.
\vspace{0.2cm}\\
{\bf Dynamical mass measurement.}
Following a previous study of the kinematics of
globular clusters in a UDG,\cite{beasley:16}
we use the tracer mass estimator (TME)
to determine the dynamical mass. This method was developed to
determine the enclosed mass from an ensemble of discrete tracers,
such as satellite galaxies or globular clusters.\cite{bahcall:81,watkins:10}
The mass within the distance of the outermost object is given by
\begin{equation}
M_{\rm TME}= \frac{C}{G}\langle v^2 r^{\alpha}\rangle r_{\rm out}^{1-\alpha},
\label{tme.eq}
\end{equation}
with $v$ the velocities
of individual clusters with respect to the mean, $r$
the projected distances of the clusters from the center of the galaxy,
$r_{\rm out}$ the distance of the furthest cluster,
and $\alpha$ the slope of the potential (with density
$\rho \propto r^{-(\alpha+2)}$).
For the case
$\alpha=1$ the potential is
similar to that of a point mass,
$\alpha=0$ corresponds
to $\rho \propto r^{-2}$ and a flat rotation curve, and for $\alpha=-1$
the density $\rho \propto r^{-1}$.
Equation \ref{tme.eq} does not take observational errors
or outliers into account; we therefore introduce the modified expression
\begin{equation}
M_{\rm TME} \approx 
\frac{C}{G} \left[ S_{\rm bi}(v_{\rm intr} r^{\alpha/2})\right]^2
r_{\rm out}^{1-\alpha}.
\label{mod.eq}
\end{equation}
Here observational errors are taken into account
by setting $v_{\rm intr} = f^{-1} (\Delta v_{\rm obs}$), with
$\Delta v_{\rm obs}$ listed in Fig.\ 2 and  $f^{-1}=\sigma_{\rm intr}/
\sigma_{\rm obs}$. $S_{\rm  bi}(x)$ denotes 
the biweight estimator of the width of the distribution.
Note that Eq.\ \ref{mod.eq}
reduces
to $M_{\rm TME} = (C/G) \sigma_{\rm intr}^2 r_{\rm out}$ for $\alpha=0$.
The constant $C$ is given by
\begin{equation}
C =
\frac{4\Gamma\left(\frac{\alpha}{2}+\frac{5}{2}\right)}{
\sqrt{\pi}\Gamma\left(\frac{\alpha}{2}+1\right)}
\,\frac{\alpha + \gamma + 1 - 2\beta}{\alpha+3 - \beta(\alpha+2)},
\end{equation}
with $\gamma$ the power-law slope of the density profile of the
clusters, $\beta = 1-\sigma_t^2/\sigma_r^2$
the Binney anisotropy parameter, 
and $\Gamma(x)$ the gamma function.
We determine the 3D density profile from a powerlaw fit
to the distribution of the spectroscopically-confirmed
globular clusters, finding $\gamma=0.9\pm 0.3$.

For an isothermal velocity dispersion profile ($\alpha=0$) and
isotropic orbits ($\beta=0$) we determine $M_{\rm TME} < 3.4
\times 10^8$\,\msun\ at 90\,\% confidence.
The results are not
very sensitive to the assumed slope of the potential or moderate
anisotropy. Changing $\alpha$ to 1 or $-1$ reduces the mass by
10\,\% or 20\,\% respectively. Tangential anisotropy with
$\sigma_t^2 = 2\sigma_r^2$ increases the mass limit to
$M_{\rm TME}<4.2\times 10^8$\,\msun; radial anisotropy with $\sigma_t^2 =0.5
\sigma_r^2$ yields $M_{\rm TME} <2.4\times 10^8$\,\msun.
We also consider errors in the density profile of the globular
clusters; for $\gamma=0.5$ the mass decreases by 20\,\% and
for $\gamma=1.5$ the mass increases by 30\,\%.
\vspace{0.2cm}\\
{\bf Robustness tests.}
As a test of the robustness of our results we consider three alternative
mass estimates. The first is the dynamical mass within the half-number
radius of the globular cluster system.\cite{wolf:10} This mass estimate
does not extend as far in radius as the TME method but is less
sensitive to the assumed level of anisotropy. For $R_{\rm gc}=3.1$\,kpc and
$\sigma_{\rm intr}<10.5$\,\kms\ we find $M_{\rm dyn}<3.2\times 10^8$\,\msun\
(see Fig.\ 4). As the halo profile is still rising at $R=3.1$\,kpc
the constraint on the halo mass is weaker than our default value, and we
find $M_{\rm halo}<8\times 10^8$\,\msun.

The second test replaces
$\sigma_{\rm bi}$ with $\sigma_{\rm rms}$, even though the rms is driven
by a single cluster (GC-98) and the  velocity distribution of the other
nine objects is inconsistent with this. The observed rms is
$\sigma_{\rm obs,rms}=14.3$\,\kms, or $\sigma_{\rm intr,rms}=12.2$\,\kms\
after taking observational
errors into account. The implied TME mass is $M_{\rm dyn} \approx
5\times 10^8$\,\msun, and the halo mass
$M_{\rm halo} \approx  3\times 10^8$\,\msun.

The third test sets the arguments against a disk aside
and assumes that the observed
velocities reflect rotation in an inclined, infinitely-thin disk.
The axis ratio of \blob\ is $b/a=0.85 \pm 0.02$, which means
that the inclination-corrected velocities are a factor of
$(\sin (\cos^{-1}\,(b/a)))^{-1} \approx 1.9$ higher than the observed ones.
Assuming an (unphysical) disk dispersion of 0\,\kms, the
inclination-corrected rotation velocity would be
$v_{\rm rot}\approx 1.4
\times 3.2 \times 1.9 
=  9_{-9}^{+14}$\,\kms, where it is assumed that the
rotation velocity is approximately 1.4 times the line-of-sight
velocity dispersion.\cite{franx:93bulge,kochanek:94} The
enclosed mass within $R=7.6$\,kpc
would be $M_{\rm disk} = 1.4_{-1.4}^{+7.6}\times 10^8$\,\msun.

For all these mass estimates the implied ratio $M_{\rm halo}/M_{\rm stars}
\lesssim 4$, the lowest ratio measured for any galaxy and two orders
of magnitude below the canonical stellar mass -- halo mass relation.
\vspace{0.2cm}\\
{\bf Data availability.}
The HST data are available in the Mikulski Archive for Space Telescopes (MAST;
http://archive.stci.edu), under program ID 14644. All other data
that support the findings of this study are available from the
corresponding author upon reasonable request.
\vspace{0.2cm}\\
{\bf Code availability.}
We have made use of standard data reduction tools in the IRAF and Python
environments, and
the publicly available codes SExtractor,\cite{bertin:96}
GALFIT,\cite{galfit} and {\tt emcee}.\cite{emcee}

\newpage
\null


\vspace{3pt}
\noindent\rule{\linewidth}{0.4pt}
\vspace{3pt}





 \newcommand{\noop}[1]{}


\begin{addendum}
 \item[Acknowledgements]
A.J.R.\ was supported by National Science Foundation grant
AST-1616710 and as a Research Corporation for Science
Advancement Cottrell Scholar.
Based on observations obtained with the W.~M.~Keck Observatory on
Mauna Kea, Hawaii.
The authors wish to recognize and acknowledge the very significant
cultural role and reverence that the summit of Mauna Kea has always
had within the indigenous Hawaiian community.  We are most fortunate
to have the opportunity to conduct observations from this mountain.
 \item[Author Contributions] P.v.D.\ led the observations,
data reduction, and analysis, and wrote the manuscript. S.D.\ visually
identified
the galaxy in the Dragonfly data and created the model galaxies to
determine the distance. Y.C.\ measured the structural parameters of
the object. A.M.\ used an automated approach to verify the
visual detections of low surface brightness galaxies in the Dragonfly
data. J.Z.\ and A.M.\  reduced the Dragonfly data. E.O.S.\ provided
the MMT image. All authors contributed to aspects of the analysis,
and to the writing of the manuscript.
 \item[Author Information] The authors declare that they have no
   competing financial interests. Correspondence and requests for
   materials should be addressed to P.v.D.~(email:
   pieter.vandokkum@yale.edu).

\end{addendum}



\appendix




\renewcommand\thefigure{Extended Data Figure \arabic{figure}}
\setcounter{figure}{0}

\begin{figure*}[htb]
   \centering
   \includegraphics[width=0.9\linewidth]{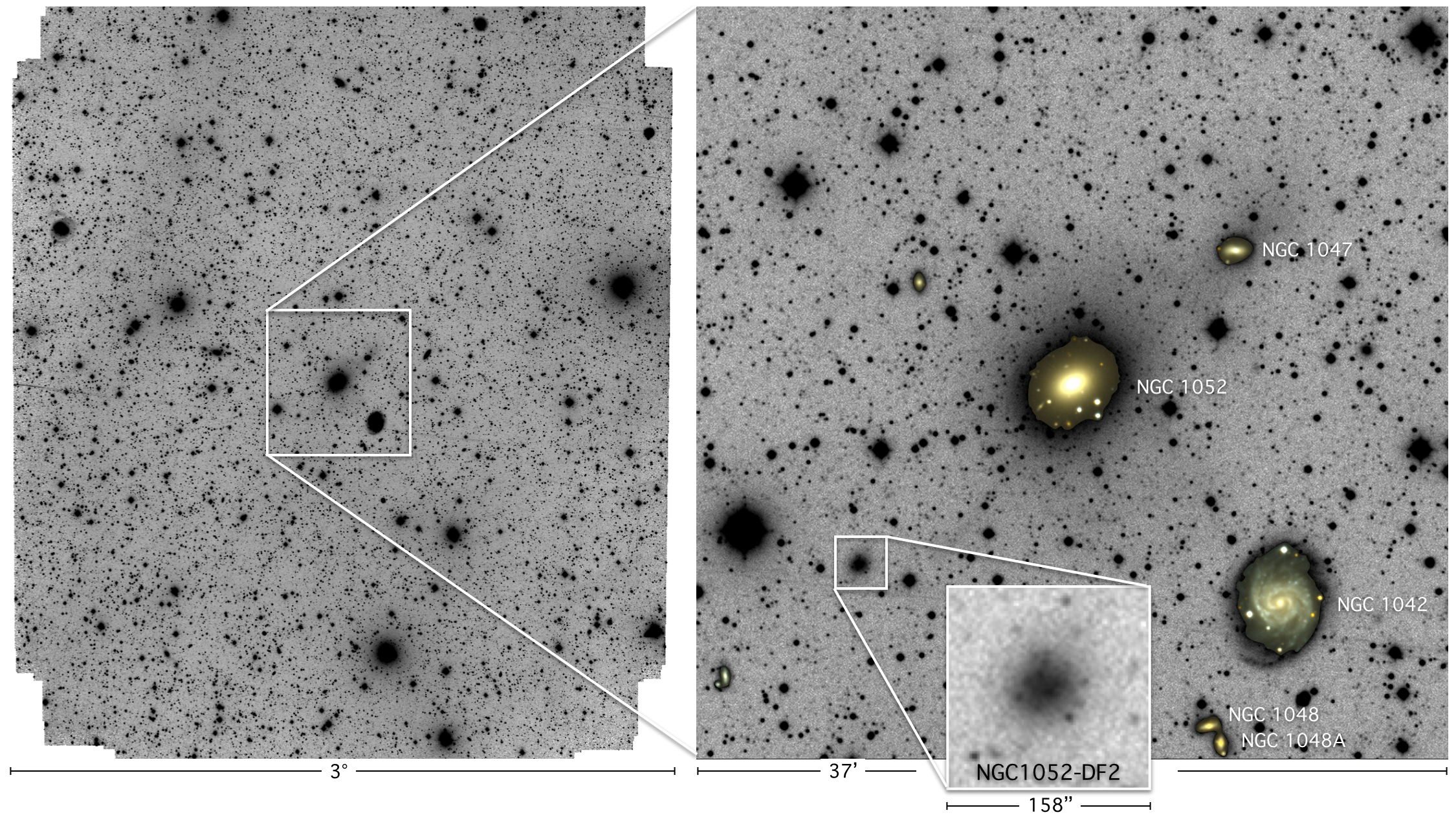}
  \vspace{-8pt}    
   \caption{ \small
\textbf{\blob\ in the Dragonfly field.} 
The full $\sim 11$\,degree$^2$ Dragonfly field
centered on NGC\,1052.  The zoom-in shows
the immediate surroundings of NGC\,1052, with \blob\ highlighted
in the inset.}
\label{dragonfly.fig}
 \vspace{-5pt}
\end{figure*}

\begin{figure*}[htb]
   \centering
   \includegraphics[width=0.45\linewidth]{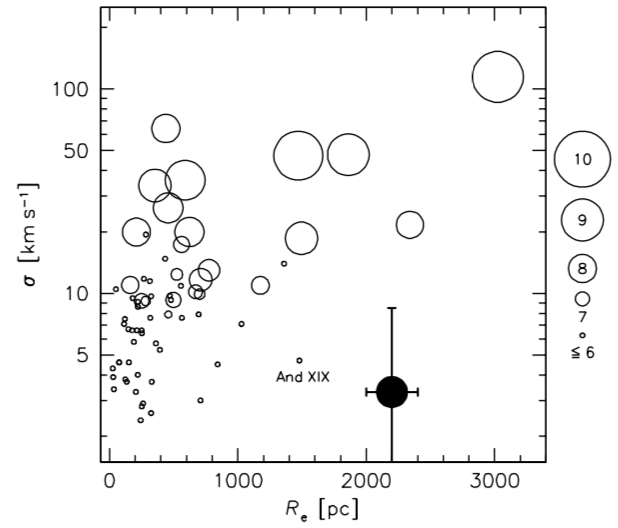}
  \vspace{-8pt}    
   \caption{ \small
\textbf{Comparison 
to Local Group galaxies.} 
Open symbols are galaxies from the Nearby Dwarf Galaxies
catalog\cite{mcconnachie:12}  and the solid symbol with errorbars
is \blob. The size of each symbol indicates the logarithm of the
stellar mass, as shown in the legend. There are no galaxies in the
Local Group that are similar to
\blob. Galaxies with a similar velocity dispersion are
a factor of $\sim 10$ smaller and have stellar
masses that are a factor of $\sim 100$ larger. The labeled
object (And\,XIX) is an
Andromeda satellite that is thought to be in the process
of tidal disruption.\cite{collins:13}
}
\label{local.fig}
 \vspace{-5pt}
\end{figure*} 

\begin{figure*}[htb]
   \centering
   \includegraphics[width=1.0\linewidth]{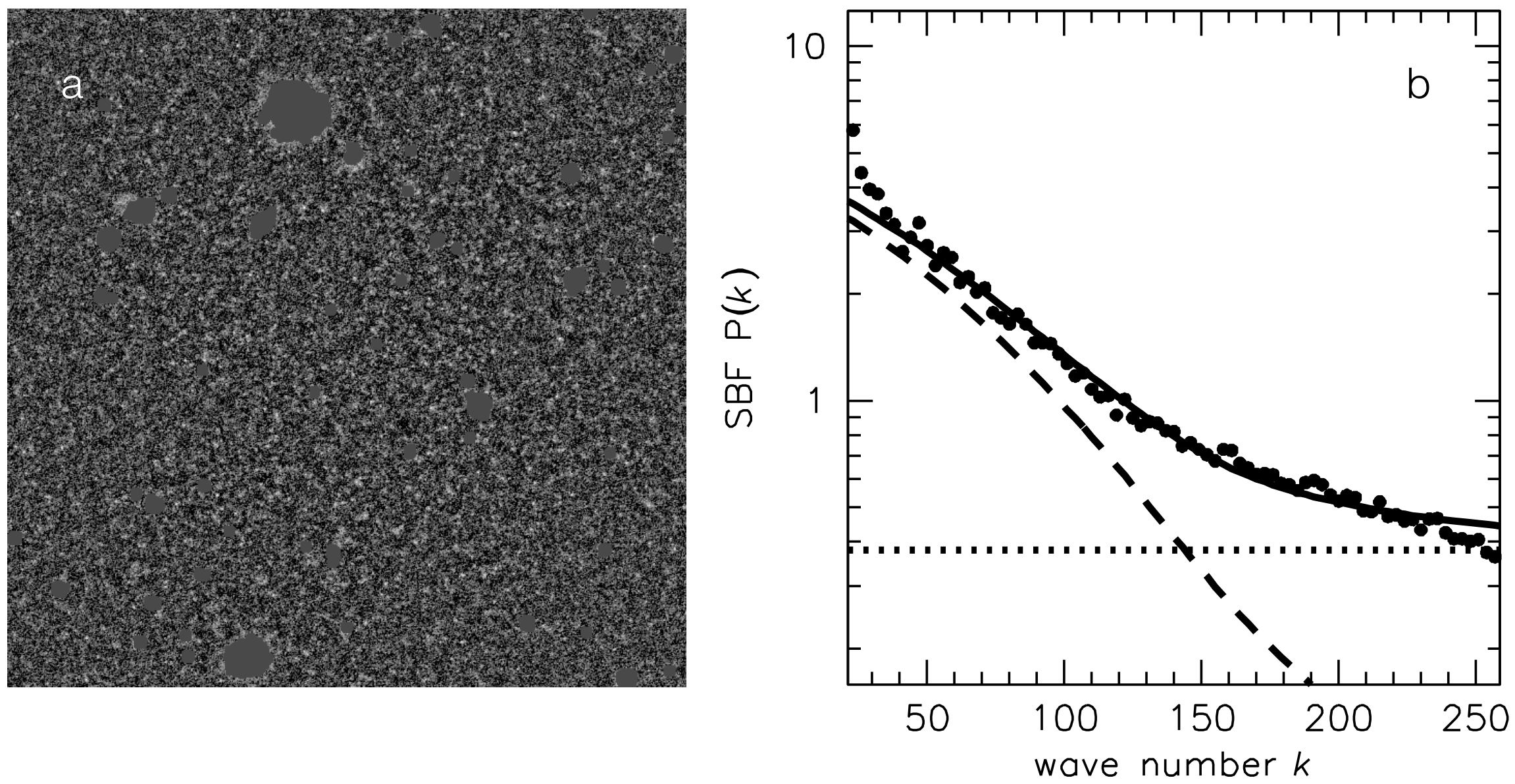}
  \vspace{-2pt}    
   \caption{ \small
\textbf{Surface brightness fluctuation analysis.}
We use the surface brightness fluctuation (SBF) signal in the
HST $I_{814}$ band to constrain the
distance to \blob. Panel a shows the galaxy after subtracting
a smooth model and masking background galaxies and globular clusters.
The image
spans  $33\arcsec \times 33\arcsec$. Panel b shows
the azimuthally-averaged power spectrum. Following previous
studies,\cite{tonry:01,mei:05,blakeslee:10} the power spectrum is
fit by a combination of a constant (dotted line)
and an expectation power spectrum $E(k)$ (dashed line).
From the normalization of $E(k)$  we find that the SBF magnitude
$\overline{m}_{814}= 29.45\pm 0.10$. The implied distance is
$D_{\rm SBF} = 19.0 \pm 1.7$\,Mpc, consistent
with the 20\,Mpc distance of the luminous elliptical galaxy 
NGC\,1052.
}
\label{sbf.fig}
 \vspace{-5pt}
\end{figure*}

\begin{figure*}[htb]
   \centering
   \includegraphics[width=0.8\linewidth]{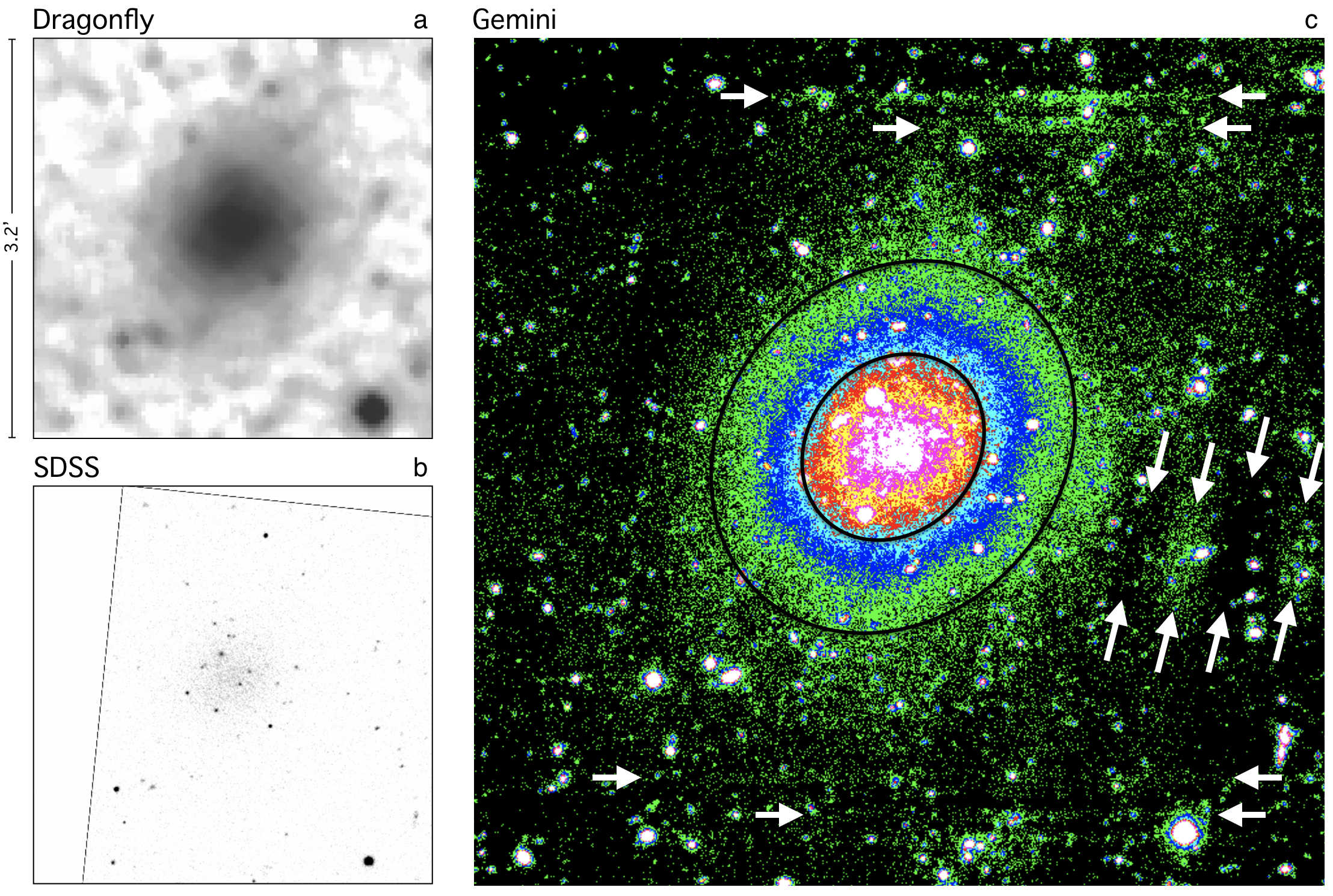}
  \vspace{-2pt}    
   \caption{ \small
\textbf{Morphological coherence.}
Panel a shows
the sum of $g$ and $r$ images taken with the Dragonfly
Telephoto Array. The image was smoothed by a $10'' \times 10''$
median filter to bring out faint emission. The lowest surface
brightness levels visible in the image are
$\approx 29$\,mag\,arcsec$^{-2}$. Panel b shows a
sum of SDSS $g$, $r$, and $i$ images. In SDSS the overdensity of
compact objects stands out.
Panel c shows the Gemini-North $i$ band image of \blob,
which
provides the best information on
the morphology of the galaxy. Black ellipses mark $R=R_e$ and
$R=2 R_e$. White arrows mark the most obvious
reduction artifacts.
The galaxy is regular out to at least $R\sim 2 R_e$, with a
well-defined center and a position angle and axis ratio that do
not vary strongly with radius.
}
\label{gemini.fig}
 \vspace{-5pt}
\end{figure*}

\begin{figure*}[htb]
   \centering
   \includegraphics[width=0.85\linewidth]{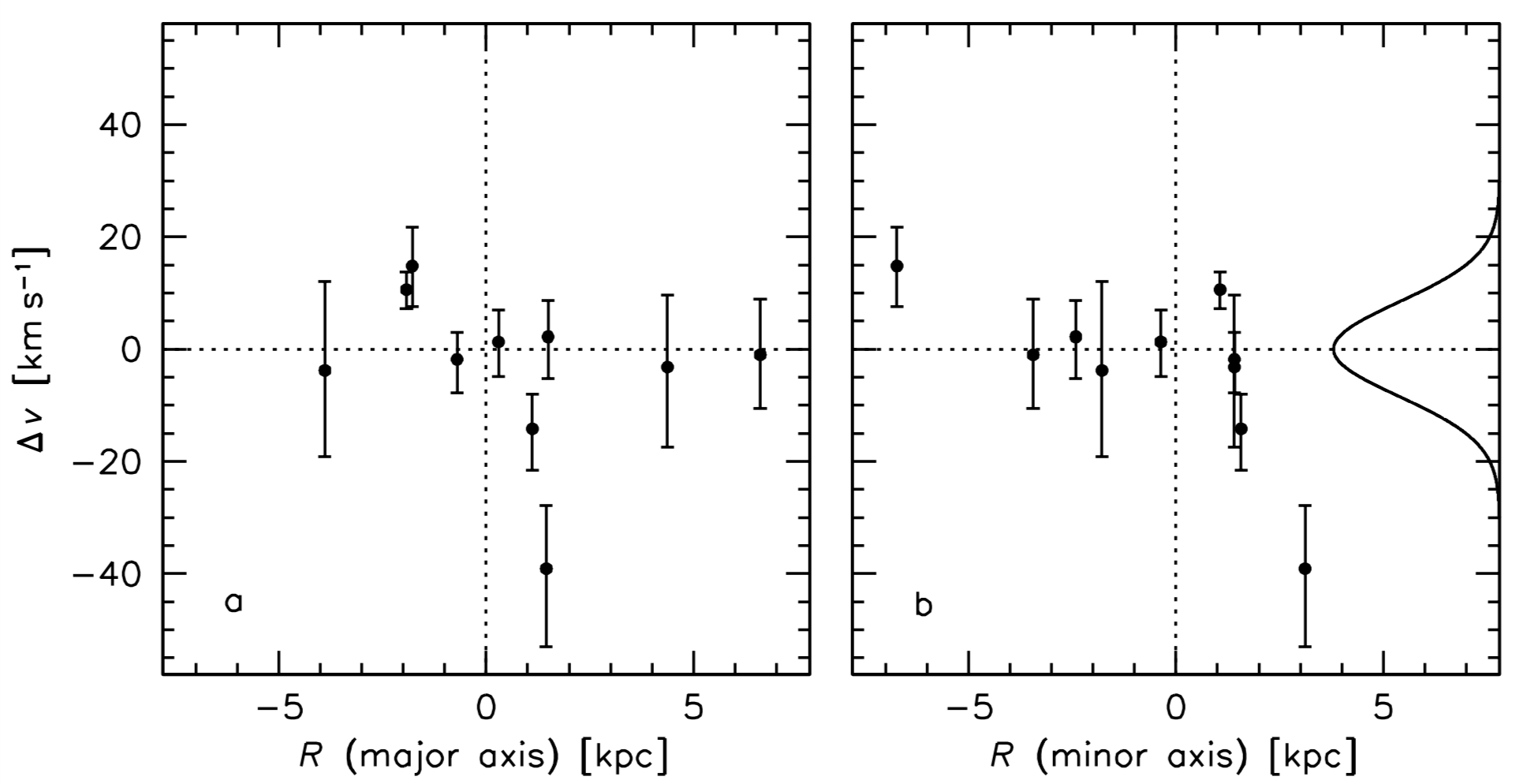}
  \vspace{-8pt}    
   \caption{ \small
\textbf{Are the globular clusters in a thin rotating
disk?} The two panels show the globular cluster
velocities as a function of projected position along the
major axis (panel a) and minor axis (panel b). Error bars are 1 s.d.\
There is no evidence for any trends. For reference, a Gaussian
with $\sigma=8.4$\,\kms\ is shown in panel b.
}
\label{rotation.fig}
 \vspace{-5pt}
\end{figure*}

%


\end{document}